\documentclass[twocolumn,
showpacs,preprintnumbers,amsmath,amssymb]{revtex4}

\usepackage{graphicx}
\usepackage{amssymb,amsmath,amsthm}
\newtheorem{Thm}{Theorem}
\newtheorem{Cor}{Corollary}
\newtheorem{Lem}[Thm]{Lemma}
\newtheorem{Prop}[Thm]{Proposition}
\theoremstyle{definition}
\newtheorem{Rmk}{Remark}

\newcommand{\bra}[1]{{\left\langle #1 \right|}}
\newcommand{\ket}[1]{{\left| #1 \right\rangle}}

\newcommand{\T}{\mbox{$\mathrm{tr}$}}
\begin{document}

\title{Distribution and dynamics of entanglement in high-dimensional quantum
systems using convex-roof extended negativity}

\author{Soojoon Lee}
\affiliation{
 Department of Mathematics and Research Institute for Basic Sciences,
 Kyung Hee University, Seoul 130-701, Korea
}
\author{Jeong San Kim}\email{jekim@ucalgary.ca}
\affiliation{
 Institute for Quantum Information Science,
 University of Calgary, Alberta T2N 1N4, Canada
}
\author{Barry C. Sanders}
\affiliation{
 Institute for Quantum Information Science,
 University of Calgary, Alberta T2N 1N4, Canada
}
\date{\today}

\begin{abstract}
We develop theories of entanglement distribution and of entanglement dynamics for qudit systems,
which incorporate previous qubit formulations.
Using convex-roof extended negativity,
we generalize previous qubit results for entanglement distribution with isotropic states
and for entanglement dynamics with the depolarizing channel,
and we establish a relation between these two types of entanglement networks.
\end{abstract}

\maketitle

Whereas shared entanglement between parties can serve as a non-local resource
with no classical counterpart, there are several practical matters we
need to consider to optimize the efficiency of entanglement usage.
One of them is the creation or {\em distribution} of entanglement.
Even though the particles share an entangled state, the state may be useless as a non-local resource in
quantum information tasks unless sufficient entanglement for the task
is shared between
desired parties to meet requirement for the task.
It is thus important to distribute entanglement
between specified parties and to have a proper way of quantifying how much
entanglement can be distributed.

Another practical matter we have to consider is the fragile nature of entanglement:
although sufficient entanglement is shared between desired parties,
this entanglement may be destroyed gradually due to interaction with the environment, which is known as {\em decoherence}.
Thus, it is also an important task to quantify or to bound the amount of entanglement
under dynamics.

For distribution of entanglement, an analytical upper bound for
the amount of remote entanglement distribution (RED) was shown
for qubit systems~\cite{GS} using {\em concurrence}~\cite{Wootters}
as a bipartite entanglement quantification. Using concurrence, entanglement evolution
including decoherence-induced non-unitary evolution
of a two-qubit state under a local channel was shown to be bounded
by the amounts of entanglement for the initial state and the
entanglement evolution of a maximally entangled state through the
channel~\cite{KDTKAB}.
Later, each of these results has been
generalized independently for higher-dimensional quantum systems in
terms of G-concurrence~\cite{Gour,TDB}.

The study of quantum entanglement in higher-dimensional
quantum systems is important and necessary for both theoretical and
practical reasons to optimize the efficiency of entanglement usage.
For example, in the applications of quantum information
theory such as quantum cryptography and quantum error correcting
code (QECC), higher-dimensional quantum systems are sometimes preferred
because of their coding efficiency and security~\cite{gjvw}.

However, generalizations of such security or efficiency proofs
from the case of qubits to the case of qudits are usually non-trivial, for instance, the
proof of a no-go theorem for the universal set of transversal
gates in QECC~\cite{cccz}.
Thus, the extension of entanglement analysis, especially the quantitative
relations of entanglement from qubit to qudit systems is a fundamental step that is
necessary to understand the whole concept of quantum entanglement.

Here we provide an explicit relation
between two different theories,
entanglement distribution and entanglement dynamics
in arbitrary dimensional quantum systems
for the isotropic states and
the generalized depolarizing channel.
Based on the Choi-Jamio{\l}kowski isomorphism between states and a quantum channel~\cite{CJiso},
we note that
the evolution of a two-qudit state through two one-sided channels corresponds to RED
by means of a joint measurement on the two qudits
consisting of each one particle of a pair of two-qudit states.
Furthermore, by using {\em convex-roof extended negativity} (CREN)~\cite{LCOK} as a bipartite
entanglement quantification, we also overcome the lack of separability criterion of G-concurrence
in the previous results~\cite{Gour,TDB},
whereas G-concurrence only detects genuine $d$-dimensional entanglement whose reduced density matrix
has full rank.

For a pure state $\ket{\phi}_{12}$ in $d\otimes d'$ quantum systems ($d'\ge d$),
its CREN is
\begin{equation}
\mathcal{N}_c(\ket{\phi}_{12}\bra{\phi}):=\frac{(\mathrm{tr}\sqrt{\rho_1})^2-1}{d-1},
\label{eq:CREN_pure}
\end{equation}
with $\rho_1=\T_2\ket{\phi}_{12}\bra{\phi}$.
For a mixed state $\rho_{12}$, its CREN is
\begin{equation}
\mathcal{N}_c(\rho_{12}):=\min \sum_{k} p_k
\mathcal{N}_c(\ket{\phi_k}_{12}\bra{\phi_k}),
\label{eq:CREN_mixed}
\end{equation}
where the minimum is taken over all possible pure-state decompositions of $\rho_{12}$
such that $\rho_{12}=\sum_k p_k \ket{\phi_k}_{12}\bra{\phi_k}$.

CREN satisfies the separability criterion for
any bipartite mixed state of arbitrary dimension, and it does not increase
under local quantum operations and classical communication (LOCC)~\cite{LCOK}.
Furthermore, CREN is relevant to the study of multipartite entanglement,
especially in characterizing the monogamy of entanglement in multi-party systems~\cite{kds}.

CREN reduces to concurrence for two-qubit systems.
In other words, CREN can be considered as a generalization
of the two-qubit concurrence, and thus the results of entanglement distribution and entanglement dynamics
for qubits systems~\cite{GS, KDTKAB} can be rephrased as the following two propositions, respectively.

\begin{Prop}\label{Prop:GS}
Assume that
Alice, Bob, and supplier Sapna perform
general LOCC on an initial four-qubit state $\rho_{12}\otimes \rho_{34}$
with outcome $\{Q_j, \sigma_{14}^j\}$
(Alice, Bob, and Sapna hold the subsystems 1, 4, and 23, respectively.);
then
\begin{equation}
\sum_{j=1}^s Q_j \mathcal{N}_c(\sigma_{14}^j)
\le\mathcal{N}_c(\rho_{12})\mathcal{N}_c(\rho_{34}).
\label{eq:REDineq2}
\end{equation}
\end{Prop}

\begin{Prop}\label{Prop:KDTKAB}
Let $\$ $ be an arbitrary one-qubit channel; then
for an initial two-qubit mixed state $\rho$,
\begin{equation}
\mathcal{N}_c\left[\left(\mathcal{I}\otimes \$\right)\left(\rho\right)\right]
\le \mathcal{N}_c\left[\left(\mathcal{I}\otimes \$\right)
\left(\ket{\phi^+}\bra{\phi^+}\right)\right]\mathcal{N}_c(\rho),
\label{eq:DoE_ineq2}
\end{equation}
where $\ket{\phi^+}=\left(\ket{00}+\ket{11}\right)/{\sqrt 2}$.
\end{Prop}

Now we show a relation between these two results for the isotropic
states and the generalized depolarizing channel in arbitrary qudit
systems. Before this, we begin with some useful terminology and notation.

Let
\begin{equation}
X=\sum_{j=0}^{d-1} \ket{j+1}\bra{j},~~
Z=\sum_{j=0}^{d-1} \omega^j\ket{j}\bra{j},
\label{eq:d_Pauli}
\end{equation}
be unitary operators on qudit systems
where $\omega=e^{2\pi i/d}$. Any operator spanned by $X$ and $Z$ is referred to as a generalized
Pauli operator~\cite{pz, gkp, bgs}.
Moreover, due to the commuting relation
\begin{equation}
ZX=\omega XZ,
\label{comm}
\end{equation}
each generalized Pauli operator can be expressed as
$X^kZ^l$ for some integers $k$ and $l$, up to constant factors.

For $0\le k, l\le d-1$, let
\begin{eqnarray}
\ket{\Psi_{k,l}}
&\equiv&\frac{1}{\sqrt{d}}\sum_{j=0}^{d-1}\omega^{jl}\ket{j,j+k}\nonumber\\
&=&Z^l\otimes X^k\ket{\Phi^+}=I\otimes X^kZ^l\ket{\Phi^+}
\label{eq:dBell_prop}
\end{eqnarray}
be the two-qudit generalized Bell states~\cite{bbcjw}, where
$\ket{\Phi^+}=\ket{\Psi_{0,0}}=\frac{1}{\sqrt{d}}\sum_{j=0}^{d-1}\ket{jj}$ is the $d$-dimensional
maximally entangled state.
The set of all generalized Bell states forms an orthonormal basis for two-qudit systems,
which is called the generalized Bell basis.

For $0\le F \le 1$,
let
\begin{align}
\$_F:~\rho\mapsto F\rho&+\frac{1-F}{d^2-1}
\left[\sum_{j,k=1}^{d-1}X^jZ^k\rho Z^{-k}X^{-j}\right.\nonumber\\
&+\left.
\sum_{j=1}^{d-1}\left(X^j\rho X^{-j} + Z^j\rho Z^{-j} \right)
\right]
\label{eq:depolarization}
\end{align}
be the generalized depolarizing channel in qudit systems with fidelity $F$,
and let
\begin{equation}
\rho_F := F\ket{\Phi^+}\bra{\Phi^+}
+ \frac{1-F}{d^2-1} \left(I\otimes I - \ket{\Phi^+}\bra{\Phi^+}\right)
\label{eq:isotropic}
\end{equation}
be a two-qudit isotropic state, then we readily have the following propositions.
\begin{Prop}\label{Prop:depolar_prop}
Any generalized Pauli operator $P$ commutes with
the depolarizing channel $\$_F$; that is,
\begin{equation}
\$_F(P\rho P^\dagger)=P\$_F(\rho)P^\dagger.
\label{eq:Pauli_depolar}
\end{equation}
\end{Prop}
\begin{Prop}
For an isotropic state $\rho_F$ and the depolarizing channel $\$_F$,
we have
\begin{equation}
(\$_F\otimes \mathcal{I})(\ket{\Phi^+}\bra{\Phi^+})=\rho_F
=(\mathcal{I}\otimes \$_F)(\ket{\Phi^+}\bra{\Phi^+}).
\label{eq:rho_F}
\end{equation}
\label{rho_F_iso}
\end{Prop}

Now, let us consider an initial four-qudit state
$\rho_{F_0}^{12}\otimes \rho_{F_1}^{34}$, which is a product of two
isotropic states $\rho_{F_0}^{12}$ and $\rho_{F_1}^{34}$ in the
subsystems $12$ and $34$ respectively with $0\le F_0, F_1\le 1$. If
the subsystem $23$ is measured in the generalized Bell basis with
the measurement outcome $|\Psi_{k,l}^{23}\rangle$ for some $k$ and
$l$, the resultant state in subsystem $14$ becomes
\begin{equation}
\sigma_{k,l}^{14}\equiv
\frac{\T_{23}\left[\left(I_{14}\otimes |\Psi_{k,l}^{23}\rangle
\langle\Psi_{k,l}^{23}|\right)\rho_{F_0}^{12}\otimes \rho_{F_1}^{34}\right]}{{Q_{k,l}}}
\label{eq:resulting_state}
\end{equation}
where
\begin{equation}
Q_{k,l}\equiv\T\left[\left(I_{14}\otimes |\Psi_{k,l}^{23}\rangle
\langle\Psi_{k,l}^{23}|\right)\rho_{F_0}^{12}\otimes \rho_{F_1}^{34}\right]
\label{resulting prob}
\end{equation}
is the probability of outcome $|\Psi_{k,l}^{23}\rangle$.
From Propositions~\ref{Prop:depolar_prop} and~\ref{rho_F_iso} together with Eq.~(\ref{eq:dBell_prop}), we have
\begin{eqnarray}
\sigma_{k,l}^{14}
&=&(\$_{F_0}\otimes \$_{F_1})(\ket{\Psi_{k,l}}\bra{\Psi_{k,l}})\nonumber\\
&=&\left(I\otimes X^kZ^l\right)
\sigma_{0,0}^{14}
\left(I\otimes Z^{-l}X^{-k}\right),
\label{eq:resulting_state2}
\end{eqnarray}
where
\begin{equation}
\sigma_{0,0}^{14}= (\$_{F_0}\otimes \$_{F_1})\left(\ket{\Phi^+}\bra{\Phi^+}\right),
\end{equation}
and hence $\sigma_{k,l}^{14}$
is equivalent to
the state $\sigma_{0,0}^{14}=(\$_{F_0}\otimes \$_{F_1})(\ket{\Phi^+}\bra{\Phi^+})$
up to local unitary operations.
Furthermore, Propositions~\ref{Prop:depolar_prop} and \ref{rho_F_iso}
readily yield
\begin{align}
(\$_{F_0}\otimes \$_{F_1})&(\ket{\Phi^+}\bra{\Phi^+})\nonumber\\
&=(\mathcal{I}\otimes \$_{F_1})(\rho_{F_0})
\nonumber \\
&=\frac{d^2{F_1}-1}{d^2-1}\rho_{F_0}+\frac{1-{F_1}}{d^2-1}I\otimes I \nonumber \\
&=\rho_{F'},
\label{eq:resulting}
\end{align}
where $F'=(d^2{F_0}{F_1}-{F_0}-{F_1}+1)/(d^2-1)$.
Thus, we obtain the following lemma.
\begin{Lem}\label{Lem:resulting_state}
For a four-qudit state $\rho_{F_0}^{12}\otimes \rho_{F_1}^{34}$,
which consists of two-qudit isotropic states $\rho_{F_0}^{12}$ and
$\rho_{F_1}^{34}$ in subsystems $12$ and $34$ respectively, if
the subsystem $23$ is measured in the generalized Bell basis, then the
resulting state $\sigma_{k,l}^{14}$ in the subsystem $14$ is equivalent
to an isotropic state $\rho_{F'}$ with
\begin{equation}
F'=\frac{(d^2{F_0}{F_1}-{F_0}-{F_1}+1)}{(d^2-1)}
\end{equation}
up to local unitary operations.
\end{Lem}

Here, we note that CREN can be analytically evaluated for the class
of isotropic states in $d\otimes d$ quantum systems~\cite{LCOK}; that
is,
\begin{equation}
\mathcal{N}_c(\rho_F)=\max\left\{\frac{dF-1}{d-1},0\right\}.
\label{eq:N_isotropic}
\end{equation}

Thus, from Lemma~\ref{Lem:resulting_state}, we can obtain the
following bound on RED by the generalized Bell measurement onto the
two-qudit subsystem 23 over the two isotropic states
$\rho_{F_0}^{12}\otimes \rho_{F_1}^{34}$.
For the case when $F' > 1/d$, it is straightforward to check that
$F_0 > 1/d$ and $F_1 > 1/d$, and thus
\begin{eqnarray}
\sum_{k,l=0}^{d-1} Q_{k,l}\mathcal{N}_c(\sigma_{k,l}^{14})
&=& \mathcal{N}_c(\rho_{F'})= \frac{dF'-1}{d-1}\nonumber\\
&=&
\frac{d^3{F_0}{F_1}-d^2-d{F_0}-d{F_1}+d+1}{(d-1)(d^2-1)} 
\nonumber \\
&\le& \frac{(d{F_0}-1)(d{F_1}-1)}{(d-1)^2}\nonumber\\
&=&
\mathcal{N}_c(\rho_{F_0}^{12})\mathcal{N}_c(\rho_{F_1}^{34}).
\label{eq:constraint1}
\end{eqnarray}
If $F'\leq 1/d$, then by Eq.~(\ref{eq:N_isotropic}), we have
$\mathcal{N}_c(\rho_{F'})=0$; hence
\begin{eqnarray}
\sum_{k,l=0}^{d-1} Q_{k,l}\mathcal{N}_c(\sigma_{k,l}^{14})
&=& \mathcal{N}_c(\rho_{F'})\nonumber\\
&\le&
\mathcal{N}_c(\rho_{F_0}^{12})\mathcal{N}_c(\rho_{F_1}^{34}).
\label{eq:constraint2}
\end{eqnarray}
We are therefore ready for the following theorem about the bound on RED for qudits
systems,
which is a generalization of the qubit case in Eq.~(\ref{eq:REDineq2})
for the isotropic states.
\begin{Thm}\label{Thm:main}
Let $F_0$ and $F_1$ be real numbers with $0\le F_0, F_1\le 1$, and
assume that the generalized Bell measurement is performed on the
subsystem 23 of the four-qudit state $\rho_{F_0}^{12}\otimes
\rho_{F_1}^{34}$. If the outcome ensemble on subsystem 14 is
$\{Q_{k,l}, \sigma_{k,l}^{14}\}$, we have
\begin{equation}
\sum_{k,l=0}^{d-1} Q_{k,l}\mathcal{N}_c(\sigma_{k,l}^{14})
\le\mathcal{N}_c(\rho_{F_0}^{12})\mathcal{N}_c(\rho_{F_1}^{34}).
\label{eq:constraint10}
\end{equation}
\end{Thm}
\begin{Rmk}
Inequality Eq.~(\ref{eq:constraint10}) is tight because it is
saturated if either ${F_0}$ or ${F_1}$ has value one, that is,
either $\rho_{F_0}^{12}$ or $\rho_{F_1}^{34}$ is a maximally
entangled state. Moreover, it also has a trivial saturation if
${F_0}$ or ${F_1}$ is less than or equal to $1/d$, as
$\mathcal{N}_c(\rho_{F'})=0=\mathcal{N}_c(\rho_{F_0})\mathcal{N}_c(\rho_{F_1})$.
These are the only cases where
Eq.~(\ref{eq:constraint10}) is saturated.
\end{Rmk}

Now, let us consider the relation between the bound on RED in
Theorem~\ref{Thm:main} and entanglement dynamics in qudit systems
~\cite{TDB}.
First, we note that, due to the Choi-Jamio{\l}kowski isomorphism and
Eq.~(\ref{eq:resulting}),
the left-hand side of Eq.~(\ref{eq:constraint10}) becomes
\begin{equation}
\mathcal{N}_c(\rho_{F'})
=\mathcal{N}_c\left[\left(\mathcal{I}\otimes \$_{F_1}\right)\left(\rho_{F_0}\right)\right],
\label{eq:LHS_constraint2}
\end{equation}
where $\$_{F_1}$ is the generalized depolarizing channel and
$\rho_{F_0}$ is the isotropic state with fidelities $F_1$ and $F_0$ respectively.
Furthermore, the right-hand side $\mathcal{N}_c(\rho_{F_0})\mathcal{N}_c(\rho_{F_1})$
of Eq.~(\ref{eq:constraint10}) can also be represented as
\begin{equation}
\mathcal{N}_c\left[\left(\mathcal{I}\otimes \$_{F_1}\right)
\left(\ket{\Phi^+}\bra{\Phi^+}\right)\right]\mathcal{N}_c(\rho_{F_0}),
\label{RHS_constraint2}
\end{equation}
due to Eq.~(\ref{eq:rho_F}) in Proposition~\ref{Prop:depolar_prop}.
Therefore, Theorem~\ref{Thm:main} can clearly be rewritten as the following corollary,
which is about the bound on
the entanglement evolution of a two-qudit isotropic state
under the action of a one-sided generalized depolarizing channel.
\begin{Cor}\label{Cor:main}
For $0\le F_0, F_1\le 1$,
we have
\begin{align}
&\mathcal{N}_c\left[\left(\mathcal{I}\otimes \$_{F_1}\right)\left(\rho_{F_0}\right)\right]\nonumber\\
&~~~~~~~~~~~~~\le \mathcal{N}_c\left[\left(\mathcal{I}\otimes \$_{F_1}\right)
\ket{\Phi^+}\bra{\Phi^+}\right]\mathcal{N}_c(\rho_{F_0}).\nonumber\\
\label{eq:constraint2}
\end{align}
\end{Cor}
\begin{Rmk}
Inequality~(\ref{eq:constraint2})
is a generalization of two-qubit entanglement dynamics in Eq.~(\ref{eq:DoE_ineq2}) to
higher-dimensional quantum systems for isotropic states and generalized depolarizing channels.
Moreover, we note that Corollary~\ref{Cor:main} also implies Theorem~\ref{Thm:main},
because both are direct consequences of
Eq.~(\ref{eq:constraint1}). In other words, the theories of RED in Theorem~\ref{Thm:main}
and entanglement dynamics in Corollary~\ref{Cor:main} are fundamentally equivalent in any qudit systems
for isotropic states and generalized depolarizing channels.
\end{Rmk}

To summarize, we have 
generalized entanglement distribution and dynamics of entanglement
into arbitrary qudit systems for isotropic states and
generalized depolarizing channels in terms of the CREN,
and established a relation between them.
Here, we note that isotropic states form a generic class of quantum states,
to which any arbitrary quantum state
can be transformed via LOCC.
Furthermore,
the generalized depolarizing channel is
the most typical case of quantum decoherence in the sense that
any quantum channel (and thus any quantum decoherence) together with
certain LOCC is equivalent to the generalized depolarizing channel.
Therefore, our result deals with
generalizations of two different theories and their explicit relation
for a generic class of quantum states and typical quantum channels
in arbitrary dimensional quantum systems.

The study of quantum entanglement, especially for higher-dimensional
quantum systems, is important and even necessary for various
applications. However, the generalization of entanglement properties
from qubits to qudits is usually nontrivial even
for the case of quantifying entanglement: still there is no
universal entanglement measure that has analytic way of evaluation,
besides qubit systems. Due to this lack of analytical evaluation,
the properties of quantum entanglement in higher-dimensional systems
are barely understood so far.

Here we have shown that CREN is a powerful entanglement measure
generalizing the quantitative bounds for both entanglement
distribution and entanglement dynamics into higher-dimensional
quantum systems. Furthermore, we have also proposed a clue
to possible unification of these different theories for arbitrary 
Hilbert-space dimensional quantum systems
by showing their explicit relation for a generic class quantum states
and quantum channels.

The authors appreciate G. Gour and T. Konrad for valuable discussions.
This work has been supported by
{\it i}CORE, MITACS and USARO. SL was also supported by Basic
Science Research Program through the National Research Foundation of
Korea (NRF) funded by the Ministry of Education, Science and
Technology (Grant No.~2009-0076578). BSC is a CIFAR Fellow.

\end{document}